\documentclass[12pt]{iopart}
\begin{document}

\title {Ultrafast non-linear optical signal from a single quantum dot: exciton and biexciton effects}
\author{A. Reyes\dag,F.J.Rodr\'{\i}guez\ddag\footnote[3] {To
whom correspondence should be addressed (frodrigu@uniandes.edu.co)}
 and L.Quiroga\ddag}

\address
{\dag Johannes Gutenberg-Universit\"at Mainz, Institut f\"ur Physik, 
ThEP D-55099, Germany}
\address
{\ddag Departamento de F\'{\i}sica, Universidad de Los Andes, A.A.  4976, Bogot\'a D.C., Colombia}

\begin{abstract}
We present results on both the intensity and phase-dynamics of the transient non-linear optical 
response 
of a single quantum dot (SQD).
 The time evolution of the Four Wave Mixing (FWM) signal on a subpicosecond time scale is dominated 
by biexciton effects. In particular, for the cross-polarized excitation case a biexciton bound 
state is found. 
In this latter case, mean-field results are shown to give a poor description of the non-linear 
optical signal at small times.
By properly treating exciton-exciton effects in a SQD, coherent oscillations in the FWM signal are 
clearly demonstrated.  These oscillations, with a period corresponding to the inverse of 
the biexciton binding energy, are correlated with the phase dynamics of the system's polarization 
giving clear signatures of non-Markovian effects in the ultrafast regime.

\end{abstract}
\maketitle
\submitto{\JPCM}
\vskip 0.1 truein

\section{Introduction}
Exciton dynamics experiments are attracting continuos interest because of
their suitability to explore dephasing effects of single exciton and multiexciton 
complexes\cite{bonadeo,gershoni}.
Furthermore, recent proposals for
 solid state quantum computing systems\cite{quiroga,sham,hawrylack,piermarochi,hohenester}, have stressed
the importance of properly controlling multiexciton coherences.
Recently, the non-classical behaviour of the
light emitted from a SQD, photon antibunching in
the fluorescence spectrum, has been observed in the ultrafast regime\cite{becher}.
Additionally, the search for single photon sources\cite{moreau} has triggered experimental interest
in transient multiexciton coherences. All of these exciton based phenomena reflect
the importance of interparticle interactions (electrons and holes) and their couplings 
to the environment, e.g, phonons. 
Both coherent and incoherent effects are experimentally detected.
 Incoherent effects are related to the system-bath interaction and from a theoretical point of
view they are usually modeled within a Markov approximation.
Coherent effects are associated with multiple particle correlations and can be observed in the subpicosecond
time scale. 
Consequently, a detailed understanding of the coherence persistence at short 
times is of most importance in nanostructure systems.

Time resolved FWM experiments give signatures of
particle-particle correlations and particle-environment interactions. 
In a typical FWM experiment, two simultaneous excitation pulses co- or cross-circularly 
polarized, with wave vectors
${\vec k_1}$ and  ${\vec k_2}$, propagate onto the sample.
 A third laser pulse, with wave vector ${\vec k_3}$, is sent time delayed at $t_2=T$. For moderate
excitation intensities, exciton-exciton interaction arises.
The time evolution of the third order polarization
$P^{(3)}(t,T)=|P^{(3)}(t,T)| e^{i\Phi(t,T)}$ describes the properties 
of the diffracted light in the ${\vec k_3}+{\vec k_2}-{\vec k_1}$ direction. 
At very short times after the excitation, particle correlations build up. 
These correlations have been observed using 
spectral interferometry techniques in bulk systems\cite{lepetit}.  
It is also possible to study the polarization phase dynamics itself, $\Phi(t,T)$.
 Moreover, other experiments for measuring $P^{(3)}$, like pump and probe schemes, 
have also been performed.
In particular in Ref.\cite{sham} a simple four level model has been used to
get the non-linear optical response ($\chi^{(3)}$) of a SQD. 
From the frequency spectrum of $\chi^{(3)}$, the possibility
of getting exciton entanglement in a SQD has been predicted. 
However, in that work exciton correlations were included within a Mean Field Approach (MFA). 

Exciton dephasing times ($T_2$) have been well characterized in bulk and quantum well systems. 
In particular, ultrafast dynamics experiments, like time-resolved speckle analysis, have been used 
 to get 
 dephasing times information in solids\cite{feuerbacher,bigot,langbein}.
On the other hand, recent experiments in bulk systems show the build up of many particle 
correlations in the femtosecond
scale of time\cite{abstreiter}. Recently, all of these dynamical phenomena have been observed for first
time in Self Assembled QDs using transient FWM spectroscopy\cite{borri} where the time integrated FWM signal oscillates
with the biexciton binding energy.  
Besides that, dephasing mechanism from Optical Longitudinal (LO) phonons has 
been observed in CdSe Bulk and quantum dots systems, showing clearly non-markovian
 effects associated to the coupling between carriers and LO phonons\cite{wogon}.
Due to its atomic like characteristics, a SQD offers the unique possibility of manipulating 
 the number of particles and their Coulomb interactions (exciton-exciton (X-X)) 
with direct consequences on decoherence processes control. Thus, rather long dephasing times
($T_2\approx$ 40 ps) have been reported\cite{kamada,bacher,zwiller}.  

 A natural question arises when  ultrafast scale of times are 
involved: what kind of quasiparticle
(free electron-hole pairs, 
exciton or bound excitons) are dominant?. 
In order to answer this question we consider the Phase Space Filling 
(PSF) effect and X-X interactions on the same footing. 
Within MFA, the FWM signal shows contributions coming from both, PSF effects 
as well as 
renormalized X-X interactions. However, this approximation fails to explain
correlations at very short times. Therefore, MFA should be improved by considering 
quantum fluctuation effects to
explain transient FWM experiments. In particular, more elaborated
theoretical approaches like the one reported in Ref.\cite{axt}, where a truncation in the hierarchy of 
polarization equations of motion
 to fifth order in the optical electric field is proposed, are shown to improve MFA results by including
four particle correlations.  Aditionally, some previous reported works\cite{axt,oestreich}
 show that in bulk 
systems a mean-field 
approach to calculate the FWM signal gives errors because the exact X-X correlations are neglected. 

In the present work, we go beyond previous limitations like few level systems and MFA. 
We use a truncation scheme similar to  one developed for bulk systems\cite{oestreich},
 in which the contribution from exciton and exciton-exciton effects can be split and treated on an
 equal footing to any desired order in the optical field\cite{axt}. However, the truncation scheme
 fails to evaluate exactly the
X-X Coulomb interaction due to the intensive computational  work involved in
its application to  bulk systems. In order
to get feasible results, this scheme has been used mapping the problem to a one-dimensional
Hubbard model\cite{oestreich}. In the present work we undertake 
this problem and we include exact X-X correlations 
 in order to  find the non-linear optical signal, intensity and phase dynamics
in a realistic model of a
 SQD in the ultrafast regime. In Section 2, we  briefly review the theoretical background
on which our results are based. Section 3 describes the main results of this work. Our conclusions
are summarized in Section 4.

\section{Theoretical Model}

Our starting point is the Hamiltonian for a system of electrons and holes in a two-dimensional
 parabolic quantum dot 
excited by external laser pulses
\begin{eqnarray}
\nonumber
H=& &\sum_{\nu,s} E_{\nu,s}^e c_{\nu,s}^{\dag}  c_{\nu,s} + 
\sum_{\nu,s} E_{\nu,s}^h h_{\nu,s}^{\dag}  h_{\nu,s}
\\
\nonumber
&+&\sum_{\nu,\sigma(s,s')}\mu_{\nu,\sigma}^{eh}{\mathcal E}(t) c_{\nu,s}^{\dag} h_{\nu,s'}^{\dag} + 
\sum_{\nu,\sigma(s,s')}\mu_{\nu,\sigma}^{he}{\mathcal E}(t)^*  h_{\nu,s'} c_{\nu,s} 
\\
\nonumber
&-&\sum_{\nu_1,\nu_2,\atop \nu_3,\nu_4}\sum_{s,s'}\langle \nu_1,\nu_2|V_{e-h}|\nu_3,\nu_4\rangle  
c_{\nu_1,s}^{\dag}  h_{\nu_2,s '}^{\dag} h_{\nu_3,s '}  c_{\nu_4,s}\\
\nonumber
&+&\frac{1}{2}\sum_{\nu_1,\nu_2,\atop \nu_3,\nu_4}\sum_{s,s'}\langle\nu_1,\nu_2 |V_{e-e}|\nu_3,\nu_4 \rangle
  c_{\nu_1,s}^{\dag}  c_{\nu_2,s '}^{\dag} c_{\nu_3,s '}  c_{\nu_4,s}
\\
&+&\frac{1}{2}\sum_{\nu_1,\nu_2,\atop \nu_3,\nu_4}\sum_{s,s'}\langle \nu_1,\nu_2|V_{h-h}|\nu_3,\nu_4\rangle
  h_{\nu_1,s}^{\dag}  h_{\nu_2,s '}^{\dag} h_{\nu_3,s '}  h_{\nu_4,s}
\label{hamilton}
\end{eqnarray}
where $c(h)_{\nu,s}^{\dag}, c(h)_{\nu,s}$ creates and destroys one electron(hole) 
in the state $\nu$ (labeled by quantum numbers $(n_{e(h)},m_{e(h)})$) with spin $s$ ($\uparrow$
, spin up, $\downarrow$, spin-down).
$V_{e-e}, V_{e-h}$ and $V_{h-h}$, denote electron-electron, electron-hole and hole-hole 
Coulomb interactions,
 respectively. Single particle energies for electrons (holes) are given by
$E_{\nu,s}^{e(h)}=(n_{e(h)}+m_{e(h)}+1)\omega_{e(h)}$, with confinement energies
$\omega_{e(h)}=1/m_{e(h)} l_{e(h)}^2$ ($\hbar =1$), where $l_{e(h)}$ is 
the parabolic confinement  length size of the electron(hole).
 The set of laser pulses used to excite the SQD is described by
the envelope amplitude $\mathcal{E}(t)$ and associated dipole moments
$\mu_{\nu,\sigma}^{eh}$ which take into account pulse polarizations.
The spins of the electron ($s$) and hole ($s'$) determine the corresponding
polarization index $\sigma$ {\it i.e.} $\sigma=\sigma(s,s')$.
By diagonalizing the time-independent part of the Hamiltonian Eq.(\ref{hamilton}) ($\mathcal{E}(t)=0$), 
the energies and wave functions for one and two excitons in a SQD are obtained, from which
the non-linear optical response is calculated.

Exciton and Biexciton wave functions are given respectively, in terms of 
 non-interacting electron-hole pairs, as
\begin{eqnarray}\label{eq:f-ondaX-XX}
\hspace{-1.5cm}
|n\rangle^{X}_{s,s'} &=& \sum_{\nu_1,\nu_2}\Psi^{n}_{\nu_1,\nu_2}
(c_{\nu_1,s}^{\dag}   h_{\nu_2,s'}^{\dag})|0\rangle\\
\hspace{-1.5cm}|m\rangle^{XX} &=& \sum_{\nu_1,\nu_2,\atop \nu_3,\nu_4}
\Psi^{m}_{{\nu_1,\nu_2},{\nu_3,\nu_4}}
(c_{\nu_1,\uparrow}^{\dag}  c_{\nu_2,\downarrow}^{\dag}\pm c_{\nu_1,\downarrow}^{\dag}  
c_{\nu_2,\uparrow}^{\dag})
(h_{\nu_3,\uparrow}^{\dag}  h_{\nu_4,\downarrow}^{\dag}\pm h_{\nu_3,\downarrow}^{\dag}  
h_{\nu_4,\uparrow}^{\dag})|0\rangle.
\nonumber
\end{eqnarray}
$|0\rangle$ is the SQD ground state and 
$\Psi^{n}_{\nu_1,\nu_2}$ and $\Psi^{m}_{{\nu_1,\nu_2},{\nu_3,\nu_4}}$ are the exciton and 
 biexciton amplitudes, respectively, which take 
properly
into account the carriers spin configuration. 
In the following formulas we will stick to the notation of Eq.(\ref{eq:f-ondaX-XX})
where the index $n$ labels an exciton state, whereas the index $m$ is reserved
for the labeling of biexciton states.

In discussing the time resolved spectra, we will be 
concerned with the situation in which the SQD is excited by very 
short laser pulses ($\delta$ functions) in resonance with the $1s$ heavy-exciton state. 
Therefore, a big number of single particle states has to be included in the
calculation of one and two exciton spectra.
 The applicability of the  $\delta$-pulses limit can be
extended to typical accoustic periods when interaction with LO phonons can be
neglected, compared with the deformation potential. However, it will be
very important in II-VI semiconductors due to this high enough polar coupling.
Our results are related with III-V systems, in which this coupling is low.
In this way, when  coherent phenomena coming from the oscillations of  
exciton or biexciton-phonon complexes are lost,  incoherent phenomena 
aquires more importance and after this time it is well known that a 
phenomenological model of decoherence, as the one we will adopt here, can be used.

We perform our theoretical calculations following a truncation scheme \cite{oestreich} of
 the Hilbert space
of one exciton and two interacting excitons.  After a long algebra it is possible to give 
analytical expressions  for the transient non-linear optical signal, which 
is proportional
to $P^{(3)}$. It can be described as the sum of three terms:  one representing the
PSF ($P^{PSF}$) contribution, another one corresponding to the MFA term ($P^{MF}$) 
and a third one related to the X-X 
contribution ($P^{F})$. This last term
 describes exact four-particle correlations and memory effects in terms of the
 so called force-force time correlation function.
All these terms can be calculated for a SQD described by the Hamiltonian Eq.(\ref{hamilton}),
giving place to expressions similar but not equal to those of Ref.\cite{oestreich} 
(the  differences are due to the confinement
geometry of the dot, see the discussion in the next section).
The resulting expressions for each one of these contributions for a SQD are given explicitly
by
\begin{equation}
\hspace{-3cm}
P_{n_1,\sigma_1}^{PSF}(t,T)e^{(i\omega_{x,n_1}+\Gamma)t}=-i\alpha_{n_1,\sigma_1}
\sum_{n_0,n_2,n_3,\atop \sigma_0,\sigma_2,\sigma_3}\alpha_{n_0,\sigma_0}\alpha_{n_2,\sigma_2}^*\alpha_{n_3,\sigma_3}^*
\Theta(T)\Theta(t) e^{(i\omega_{x,n_0}-\Gamma)T} 
C_{n_0,\sigma_0;n_1,\sigma_1}^{n_2,\sigma_2;n_3,\sigma_3} 
\end{equation}
where $\alpha_{n,\sigma}$ is related to the exciton wave function at zero relative distance, 
$C_{n_0,\sigma_0;n_1,\sigma_1}^{n_2,\sigma_2;n_3,\sigma_3}$
is the PSF parameter (for notation details, see Ref.\cite{oestreich}) and 
$\Theta(x)$ denotes the step function ($\Theta(0)$=1). 
The MFA contribution can be written as
\begin{eqnarray}
\hspace{-1.5cm}
\nonumber
P_{n_1,\sigma_1}^{MF}(t,T) e^{(i\omega_{x,n_1}+\Gamma)t}&=&
                                -i\alpha_{n_1,\sigma_1}\sum_
                                        {n_0,n_2,n_3,\atop \sigma_0,\sigma_2,\sigma_3} 
 \frac{
\alpha_{n_0,\sigma_0}\alpha_{n2,\sigma_2}^*\alpha_{n3,\sigma_3}^*
 e^{(i\omega_{x,n_0}-\Gamma)T}
\beta_{n_0,\sigma_0;n_1,\sigma_1}^{n_2,\sigma_2;n_3,\sigma_3}   }
      {\omega_{x,n_0}+\omega_{x,n_1}-\omega_{x,n_2}-\omega_{x,n_3}+2i\Gamma }\times\\
& &
\times\left [\Theta(-T)\Theta(t+T) A_x(t,T)+\Theta(T)\Theta(t) A_x(t,0) \right ]
\label{meanfield}
\end{eqnarray}
where
\begin{equation}
\beta_{n_0,\sigma_0;n_1,\sigma_1}^{n_2,\sigma_2;n_3,\sigma_3} 
=\sum_{m}
(\omega_{xx,m}-\omega_{x,n_0}-\omega_{x,n_1} )B_{n_0,n_1}^{\sigma_0,\sigma_1}(m)
                                              B_{n_2,n_3}^{\sigma_2,\sigma_3}(m)
\label{beta}
\end{equation}
%
%
and
\begin{equation}
\hspace{-1cm}
A_x(t,T)=e^{(-2\Gamma+i(\omega_{x,n_0}+\omega_{x,n_1}-\omega_{x,n_2}-\omega_{x,n_3}))t}-
e^{(2\Gamma-i(\omega_{x,n_0}+\omega_{x,n_1}-\omega_{x,n_2}-\omega_{x,n_3}))T}.
\label{Ax}
\end{equation}
Similarly, the X-X contribution is
given by
\begin{eqnarray}
P_{n_1,\sigma_1}^{F}(t,T)e^{(i\omega_{x,n_1}+\Gamma)t}= i\alpha_{n_1,\sigma_1}\bigg(\sum_{m}P_{n_1}(m)\bigg)
  -P_{n_1,\sigma_1}^{MF}(t,T)e^{(i\omega_{x,n_1}+\Gamma)t}
\label{xx}
\end{eqnarray}
where
\begin{eqnarray}
\hspace{-2.00cm}
\nonumber
P_{n_1}(m)=\sum_{n_0,n_2,n_3,\atop \sigma_0,\sigma_2,\sigma_3}
\alpha_{n_0,\sigma_0}\alpha_{n_2,\sigma_2}^*\alpha_{n_3,\sigma_3}^*
B_{n_0,n_1}^{\sigma_0,\sigma_1}(m)B_{n_2,n_3}^{\sigma_2,\sigma_3}(m)
\frac{e^{(i\omega_{x,n_0}-\Gamma)T}(\omega_{xx,m}-\omega_{x,n_0}-\omega_{x,n_1})}{\omega_{xx,m}-\omega_{x,n_0}-\omega_{x,n_1}-i\Gamma_{xx}}\times
\nonumber
\\
\times\left [(\Theta(-T)\Theta(t+T) A_{xx}(t,T)+\Theta(T)\Theta(t)A_{xx}(t,0)\right ] 
\end{eqnarray}
with 
\begin{equation}
A_{xx}(t,T)=e^{(-\Gamma_{xx}+i(\omega_{x,n_0}+\omega_{x,n_1}-\omega_{xx,m}))t}-
e^{(\Gamma_{xx}-i(\omega_{x,n_0}+\omega_{x,n_1}-\omega_{xx,m}))T}.
\label{Axx}
\end{equation}
In Eqs.(\ref{beta}) and (\ref{xx}), $m$ runs over all (bound and unbound) biexciton states.
Bound biexcitonic states are those for which its energy lies below two times
the ground state energy of an exciton. Aditionally, 
$\beta_{n_0,\sigma_0;n_1,\sigma_1}^{n_2,\sigma_2;n_3,\sigma_3}$ and $P_{n_1,\sigma_1}(m)$ 
take into account the exciton and biexciton
 transition weights on the
signal produced by the $n_1$ exciton state. 
The exciton transition weights $B_{n_i,n_j}^{\sigma_i,\sigma_j}(m)$ are defined to be the matrix
elements  
$\langle 0 |{\hat B}_{n_i,\sigma_i}{\hat B}_{n_j,\sigma_j}|m\rangle^{XX}$ 
of the product of
two $n^{th}$ exciton destruction operators $\hat B_{n,\sigma}$
between the exciton ground state $\langle 0|$ and a
biexciton state $|m\rangle^{XX}$.
 Finally, all we need is: exciton, $\omega_{x,n}$, and biexciton, $\omega_{xx,m}$, energies and 
their respective wave functions, which can be numerically obtained to any desired precision in the SQD
case. From these ingredients, the exciton transition weights 
can be calculated as $B_{n_0,n_1}(m)=\langle 0 |{\hat B}_{n_0}{\hat B}_{n_1}|m\rangle^{XX}$ 
where the exciton destruction operator is
$\hat B_{n,\sigma}=\sum_{\nu_1,\nu_2}\Psi^n_{\nu_1,\nu_2}c_{\nu_1,s}h_{\nu_2,s'}$
($\sigma=\sigma(s,s')$)  and 
$\Psi^n_{\nu_1,\nu_2}$ is the  $n$-th exciton amplitude (cf. Eq.(\ref{eq:f-ondaX-XX})).
One of the main results of our paper
is that, in contrast to an extended system, for QDs the term $P_{n_1,\sigma_1}^{MF}$ 
does not vanish, \emph{even} in the case of cross-polarized (CP) excitation. If the
dot confinement energies are reduced,  in the CP case $P_{n_1,\sigma_1}^{MF}$ decreases in
magnitude, approaching zero in the limit of zero confinement
(extended system). This can be seen by direct evaluation of the 
parameter $\beta$ in  Eq.(\ref{beta}), that  gives
 information on the strength of the signal as predicted by MFA. Although, the
$\beta$ term contains information about the renormalized biexciton energy, this term
is not responsible for oscillations and only contributes with an spectral
weight given by the $B_{n_i,n_j}^{\sigma_i,\sigma_j}$ changing the intensity of
the FWM signal.

The total dynamics of excitons and biexcitons comprises the MF and the
exact X-X contributions. The temporal dynamics dependence of $P^{MF}$ (apart from 
an exponential damping) is contained in Eq.(\ref{Ax}) where only
contributions from differences of exciton energy levels appear, which
are negligible in the case of resonant excitation. On the other hand, the 
temporal dependence of the X-X contribution is dominated by energy
differences between two exciton states and one biexciton state, e.g.
Eq.(\ref{Axx}). In particular, in the CP case, a bound biexcitonic 
state is present, causing oscillations in the signal with 
a frequency corresponding to the energy of the bound state. 
 
Exciton and biexciton dephasing rates associated to non-radiative mechanisms 
are described (phenomenologically) by $\Gamma$ and $\Gamma_{xx}$, respectively. 
In the following we will concentrate only in the lowest exciton state,
 i.e. $n_1=n_2=n_3=n_4=1s$ with
$\Gamma_{xx}=2\Gamma$. 
In our model, the dephasing time $T_2$ is defined as the inverse of   $\Gamma_{xx}$.

\section {Results and discussion}
In order to calculate the FWM signal for a 
typical self-assambled SQD\cite{raymond,wojs},  
we consider a $In_{0.5}Ga_{0.5}As$ quantum dot with
 $\epsilon=12.5$, $m_e^*=0.068$, $m_h^*=0.2$, $\omega_e=20$ meV and $\omega_h=5$ meV. To guarantee
a good convergence for the spectrum of zero total angular
momentum excitons and biexcitons, we diagonalize a 72x72 Hamiltonian matrix 
 for excitons and a
3250x3250 Hamiltonian matrix for biexcitons. The exciton spin is determined by the
circular polarization of the excitation pulses. Further in this work we discuss only the
cross polarization excitation case. 
With this set of parameters the biexciton binding energy is -1.27 meV

As discussed in Section 2, the FWM signal clearly depends on $T$, on the 
exciton and biexciton energies as well as on the transition spectral weights. 
Excitation pulses create coherent excitons whose dephasing time ($T_2$) can be monitored by the delayed
pulse. As the time delay increases, the FWM signal decreases from which $T_2$ can be obtained.
With these ingredients,
and considering that more than one electronic excitation can be induced by the laser 
and four particle correlations
occur for times lesser than $T_2/4$, we expect that interference
phenomena appear in the ultrafast time scale. 
Therefore, the coherent FWM signal and the phase dynamics give clear signatures of exciton-exciton
interactions in the ultrafast regime.
In the following we will not be concerned with $T_2$ determination,
which is to be taken as a free parameter. Pulses with zero time delay, $T=0$, will be
considered.

In the weak intensity regime both PSF and X-X effects determine the nonlinear optical properties. 
We will concentrate on these two contributions.  PSF contributions
 to the FWM signal are only important
in the  parallel spin exciton case, where X-X effects are negligible, 
because a bound biexciton  state does not exist at all. 
By contrast, for the cross polarization situation we are interested in, PSF effects 
are vanishing while X-X effects become dominant. 
It is important to remark that in extended and isotropic systems the MFA gives a zero contribution
to the FWM signal, because only excited excitons with zero center of mass momentum 
contribute, i.e.,($\vec q = 0$)\cite{oestreich}.
However, in SQD this condition can be relaxed due to the 
fact that the parabolic potential absorbs the incident 
momentum and excitons with center of mass momentum different from zero can be excited.
This point  becomes particularly clear by considering the mean field
parameter $\beta$ defined in Eq.(\ref{beta}). Whereas for extended systems this parameter
is exactly zero \cite{oestreich}, for a SQD it becomes a function
of the confinement energies $\omega_{e(h)}$. The dependence of $\beta$ 
on $\omega_e$ is depicted in Fig. (1).  We consider a constant relationship 
between  electron and hole confinement lengths, i.e, $l_e=l_h$. It can be seen 
that in the limit of small confinement, $i.e.$  pure two-dimensional systems ($\omega_{e,h}$=0), 
$\beta$ tends to zero, as it should 
be. This means that  the usual condition of ($\vec q \neq 0$) for the carriers Coulomb
interaction in extended systems is broken in confined systems.
We have tested our results taking a huge set of excitonic states and we found that higher
energy states give a  negligible contribution to $\beta$.

The phase-dynamics
and the non-linear optical signal are calculated for different decay rates. 
In order to better clarify, why MFA is not adequate to describe non-linear optical signals at small times, we plot 
in Fig. 2, separately, the amount of particle correlation contribution to the FWM signal
as described within MFA (Eq.\ref{meanfield})
 and exact X-X effects (Eq.\ref{xx}), 
which include 
memory (non-Markovian) effects, for $\Gamma_{xx}=0.125$ meV. 
 The usual and simplest approximation for this last term
yields to the Markov approximation and MFA results.
 By contrast, in this work we have performed
a full numerical evaluation of X-X effects by summing over all biexciton states. Hence, 
we plot the absolute values of both,
the real part of $P_{n_1,\sigma_1}^{MF}(t,T) e^{(i\omega_{x,n_1}+\Gamma)t}$ (MFA term) and 
the real part of $P_{n_1,\sigma_1}^{F}(t,T) e^{(i\omega_{x,n_1}+\Gamma)t}$ (X-X term).
%
The MFA result grows monotonically with $t$, whereas the X-X result 
shows beating oscillations with a period corresponding to $T_{xx}=2\pi\omega_{xx,opp}^{-1}\approx 5 ps$, 
the biexciton binding energy. 
As can be seen, non-Markovian effects arising from exciton-exciton correlations 
are the main source of discrepancies between
MFA results and the exact ones. This means that ultrafast spectroscopy in SQDs should register
memory effects associated to four particle correlations.
 
In the inset of Fig. 2, we show the spectral weight $B(m)$ as a function
of biexciton energy levels (the zero of energy corresponds to twice the single $1s$-heavy-exciton energy). 
We have only plotted those levels that have total angular momentum $L_z=0$. The bound
biexciton level is clearly seen at negative energies. 
Although, unbound biexciton states have non-negligible spectral weights the bound biexciton
still has a comparable weight. 
As can be seen in Fig. 2  MFA fails to describe properly four particle interactions at
very short times, where particle-particle correlations are important. 
This term does not oscillates and only average the exact result. Besides that, the 
contribution in Eq.\ref{xx}, takes into account a higher degree of correlation going
beyond MF and gives oscillations at very short times. This oscillations provide information about
strong correlations between excitons.
Thus, the bound biexciton dynamics is directly linked to 
non-Markovian effects.
By contrast, in the long time limit ($t > T_2$), memory effects 
are lost and
MFA results approach to the real part of the first term in Eq.\ref{xx}. 
The FWM signal becomes dominated by non-radiative dephasing 
effects, producing in this way a simple exponential decay. 

In order to assess how the main features described above are reflected on the optical dynamical 
characteristics, we
plot the non-linear FWM intensity $|P^{(3)}(t,0)|$ in Figs. 3(a) and 3(c), 
and the dynamical phase $\Phi(t,0)$ in Figs. 3(b) and 3(d), for dephasing rates,
 $\Gamma_{xx}=T_2^{-1}=$0.125 meV 
and 0.5 meV, respectively.
To better illustrate the results, a comparison is made between MFA and X-X results on the same plots.
For a low dephasing rate ($T_2\approx$ 8 ps), strong oscillations are clearly seen in Figs. 3(a) and (b) for 
$t < T_2$. 
The X-X interaction, induces a $\pi$ phase
shift with respect to the incident electric field. The phase of the $P^{(3)}$ signal oscillates
during a typical time $T_2$. FWM signal shows a strong peak at $t\approx T_2/4$ whereas it should have a
 maximum at $t\approx T_2$ from MFA. 
For high dephasing rates ($T_2\approx$ 2 ps), Figs. 3(c) and (d), the attenuation of both the intensity
and phase of the third order optical signal is evident. At this
time scale, the $T_2$ dephasing time is shorter than the time for which coherent effects could be seen. 
It is worthnoting that while the phase shows a similar starting behaviour as compared with
the low dephasing case, the FWM signal intensity is drastically reduced by roughly a factor of five 
(notice the change of vertical scale).

Most importantly, the phase dynamics exhibits novel features at small times when exciton-exciton
correlations are considered. By contrast, MFA results yield to a real $P^{(3)}$ for any time. Therefore,
no phase dynamics is observed within MFA. 
In particular, X-X effects present a correlation between the first FWM signal minimum and 
a $\pi$ jump in its phase.  
For long times the phase goes to a constant value. However, after the 
extinction of the FWM intensity, there is no longer a clear physical meaning for this phase.
Hence, oscillations in the phase dynamics, with a beating frequency controlled by the biexciton binding energy,
could bring important information about exciton correlations in SQDs. 
In contrast with higher dimensional systems (quantum wells and bulk samples), a SQD provides us
 with an adequate system where exciton and biexciton discrete energy spectra can be
tailored by changing the
confinement potential. In this way, optimal conditions to enhance phase memory could be explored.

\section{Conclusions}
In summary, we have studied the non-linear optical response from multiexciton complexes in a SQD.
 For cross-polarized excitation, the creation of bound biexcitons is possible. They dominate the
 non-linear 
polarization dynamics in the low density regime.
 The phase dynamics and the FWM signal in the ultrafast scale of time have been obtained.
By including all the numerically determined biexciton states, exact exciton-exciton correlations are 
evaluated allowing to go beyond simple
MFA results. In particular we found that the phase dynamics and the FWM intensity oscillate with a period 
which coincides with the inverse of bound biexciton frequency.
We demonstrated clearly that 
the usual Mean Field theory, which is closely related to the Markovian approximation, does not describe
correctly exciton correlations at very small times. However,
our results show that at long times,  MFA results for both intensity and phase dynamics,
are close to those produced by a more realistic calculation.

\ack{Acknowledgents}
 FJR and LQ  acknowledge partial support from COLCIENCIAS (Colombia) project
No.1204-05-10326/11408, Banco de la Rep\'ublica (Colombia) and from the EPSRC-DTI LINK project (UK).

\newpage
\centerline{\bf Figure Captions}

\bigskip
\noindent Fig. 1:  MFA contribution ($\beta$ term) as a function of the electron and hole
confinement frequencies, for $\Gamma_{xx}$=0.125 meV.
\bigskip

\noindent Fig. 2:  Absolute values of the MFA contribution (real part of 
$P_{n_1}^{MF}(t,T) e^{(i\omega_{x,n_1}+\Gamma)t}$)
 and X-X contribution ($P_{n_1}^{F}(t,T) e^{(i\omega_{x,n_1}+\Gamma)t}$) to the FWM
signal, for $\Gamma_{xx}$=0.125 meV.
Inset:  Spectral weight of the biexciton wave function. Bound biexciton energy=-1.27 meV. The long-dashed line denotes the zero energy.
\bigskip

\noindent Fig. 3: $\Gamma_{xx}$=0.125 meV: 
(a) Time resolved FWM signal; (b) Phase dynamics. $\Gamma_{xx}$=0.5 meV: (c) Time resolved 
FWM signal; (d) Phase dynamics.


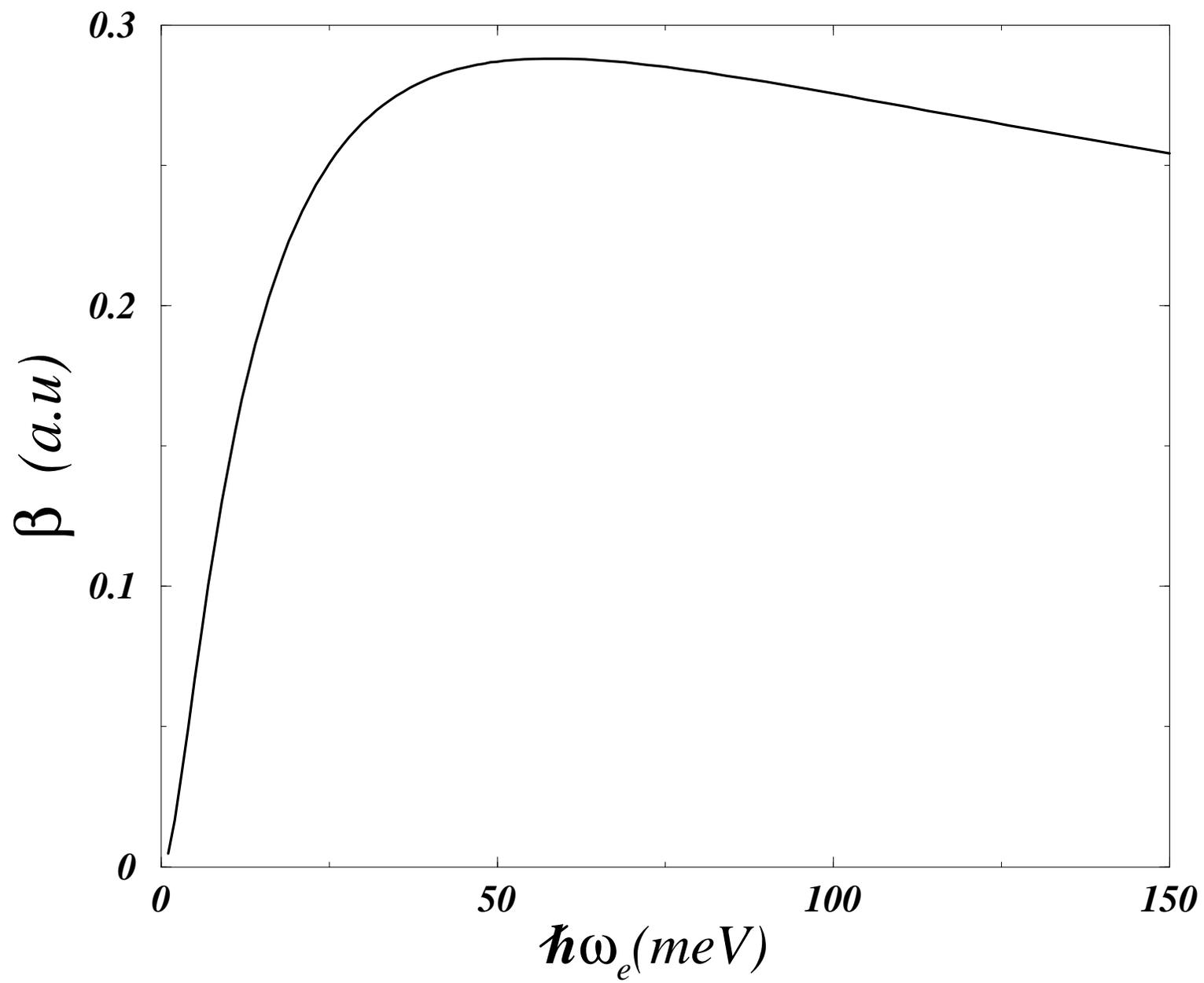
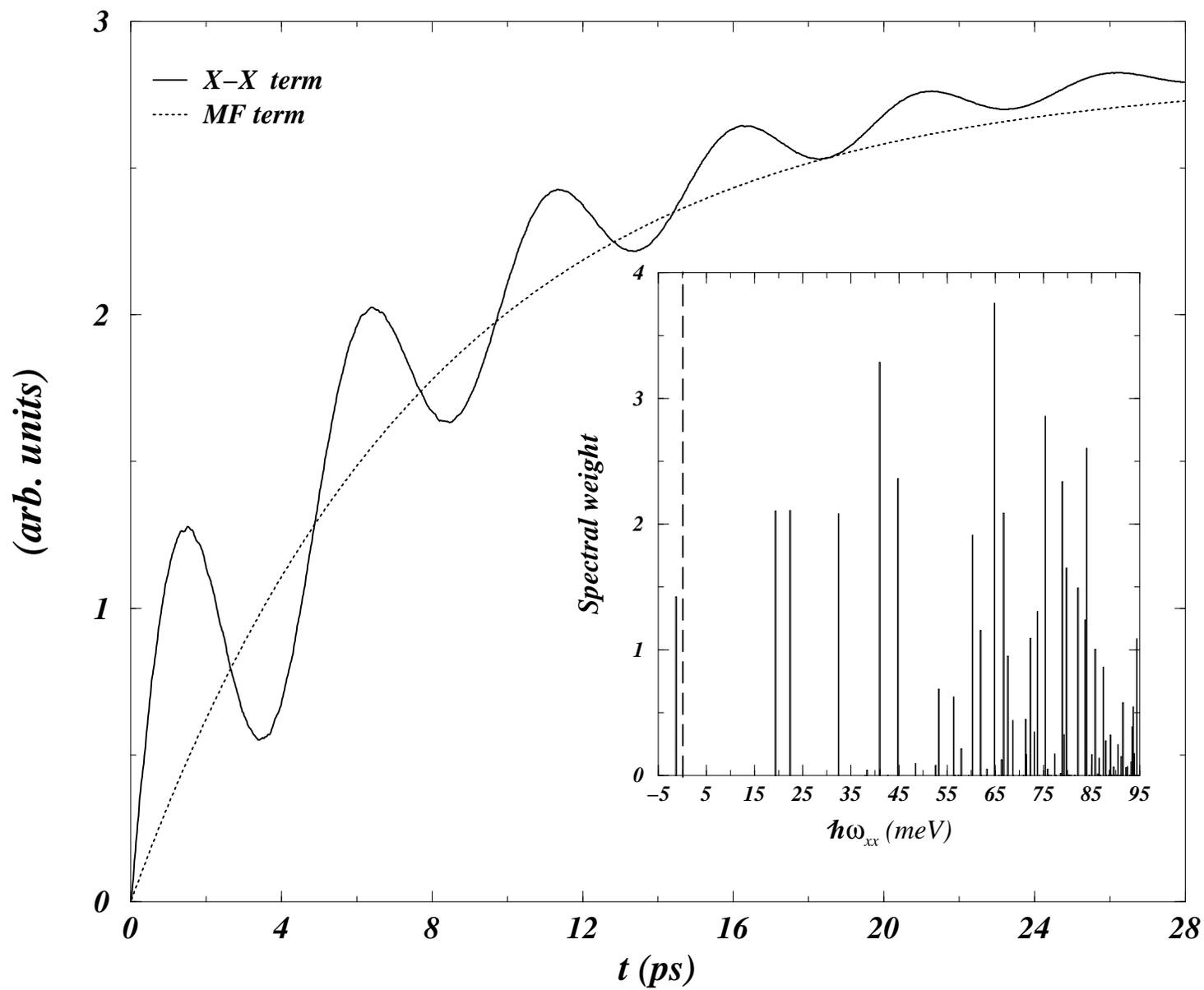
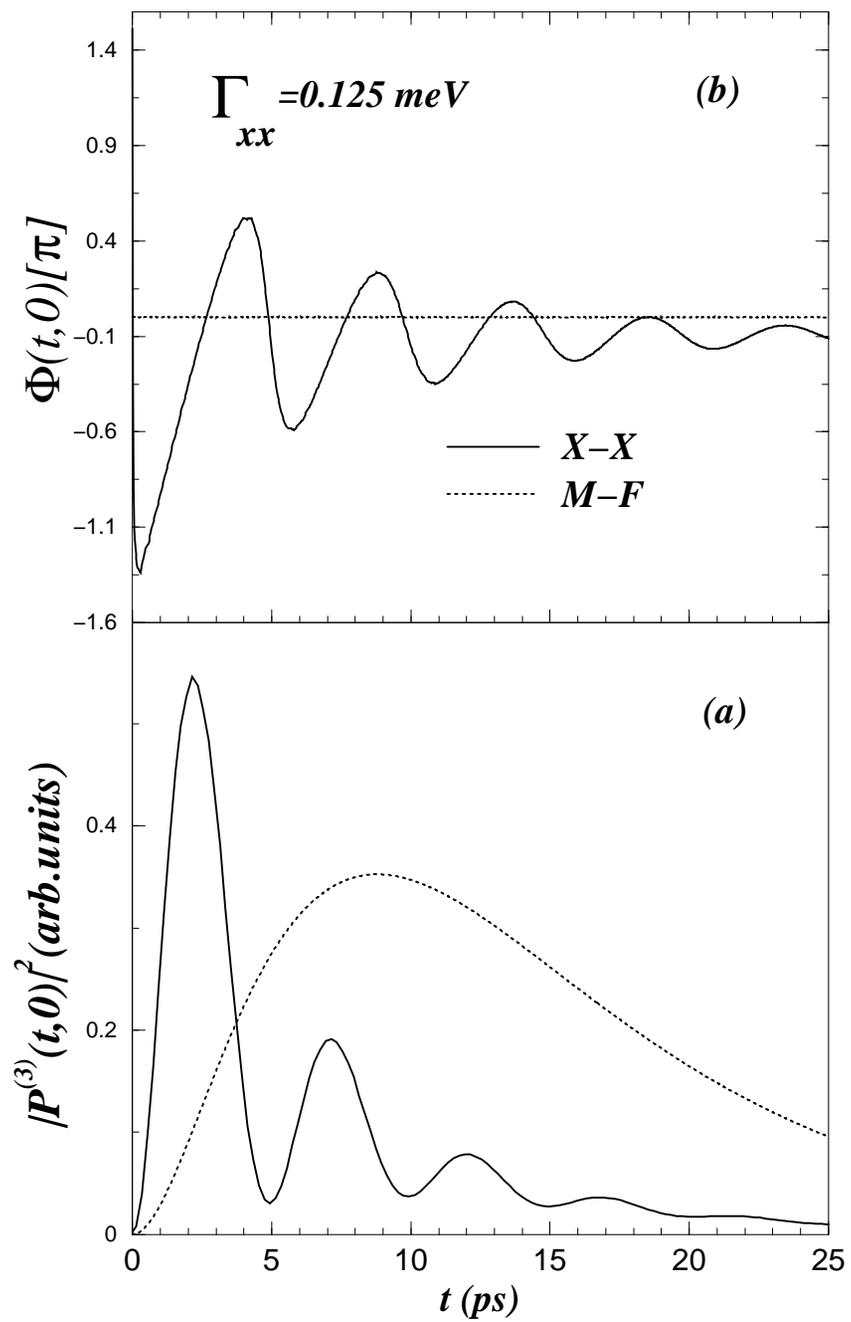
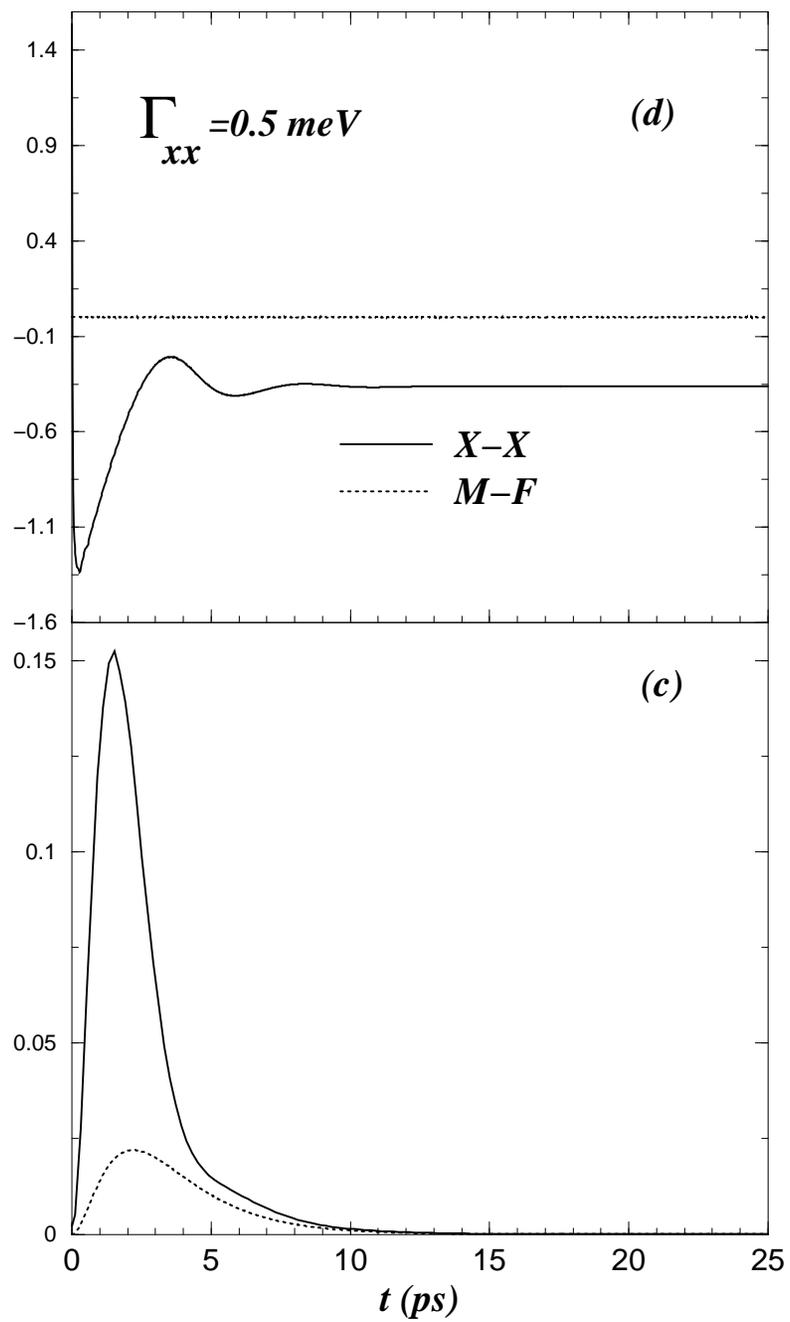
\end{document}